# Kinetic proofreading at single molecular level: Aminoacylation of $tRNA^{Ile}$ and the role of water as an editor

**Mantu Santra and Biman Bagchi***

**Solid State and Structural Chemistry Unit,**

**Indian Institute of Science,**

**Bangalore - 560012, India.**

***E-mail: bbagchi@sscu.iisc.ernet.in**

## Abstract

Proofreading/editing in protein synthesis is essential for accurate translation of information from the genetic code. In this article we present a theoretical investigation of efficiency of a kinetic proofreading mechanism that employs *hydrolysis of the wrong substrate as the discriminatory step* in enzyme catalytic reactions. We consider aminoacylation of $tRNA^{Ile}$ which is a crucial step in protein synthesis and for which experimental results are now available. We present an augmented kinetic scheme and then employ methods of stochastic simulation algorithm to obtain time dependent concentrations of different substances involved in the reaction and their rates of formation. We obtain the rates of product formation and ATP hydrolysis for both correct and wrong substrates (isoleucine and valine in our case), in single molecular enzyme as well as ensemble enzyme kinetics. The present theoretical scheme correctly reproduces (i) the amplitude of the discrimination factor in the overall rates between isoleucine and valine which is obtained as $(1.8\times10^2).(4.33\times10^2) = 7.8\times10^4$, (ii) the rates of ATP hydrolysis for both Ile and Val at different substrate concentrations in the aminoacylation of $tRNA^{Ile}$. The present study shows a non-michaelis type dependence of rate of reaction on $tRNA^{Ile}$ concentration in case of valine. The editing in steady state is found to be independent of amino acid concentration. Interestingly, the computed ATP hydrolysis rate for valine at high substrate concentration is same as the rate of formation of Ile-$tRNA^{Ile}$ whereas at intermediate substrate concentration the ATP hydrolysis rate is relatively low.

## I. Introduction

Kinetic proofreading is the theory proposed to rationalize the known lack of errors in biological synthesis. In biochemical reactions, enzymes not only enhance the rate of reaction, but also selectively choose the correct substrate leading to the desired product. Many biological processes, like protein synthesis or DNA replication, exhibit high specificity towards the selection of the correct substrates in presence of many other structurally or chemically analogous substrates [1]. Due to the similar binding energy of both the right and wrong substrates and the size/shape analogue to the enzyme, the error rate (the ratio of the rate of wrong product formation to that of the desired product formation) is expected to be high. To the contrary, the error rate is extremely low in selection of amino acid in protein synthesis ($10^{-4}$) [2] and DNA replication ($10^{-9}$) [3-5]. The molecular reason for such high selectivity is still not fully understood from a quantitative theory. This important problem has remained a debated subject for several decades, with the original Hopfield formulation of repeated activation found inadequate in several biosyntheses [6]. Recent experimental studies in several enzyme catalytic reactions reveal that the decomposition of the intermediates occurs through hydrolysis reaction [7, 8]. Several alternative editing mechanisms have been proposed and found to be satisfactory in different cases, outlining the fact that more than one mechanism could be operating [9]. One of these mechanisms, proposed first by Fersht, employs hydrolysis of the wrong substrate as the main discriminatory step.

In a previous study, we investigated the catalytic process of aminoacylation of tRNA. We showed that for class I synthetases, the steady state and single molecular events provide somewhat different kinetics [10]. We presented an augmented kinetic scheme that included the role of water and employed the versatile technique of first passage time distribution to obtain time-dependent rate. The theory explained discrepancy between single turn over and steady state rate for both class I and class II enzymes [10].

In this work we study the kinetic proofreading of class I enzyme. We studied the relative merits of two scheme proposed by Hopfield and Fersht and elucidate the rate. In order to quantify kinetic proofreading we investigated with available data the extent of correct and wrong product formation.

In a pioneering study, Fersht proposed that the most important discriminating step in kinetic proofreading could be the hydrolysis of high energy enzyme-substrate complex. Thus, a simple chemical reaction with water plays a critical role. However, a detailed quantitative analysis of the Fersht scheme has, to the best of our knowledge, not been carried out.

Here we present such an analysis. We propose a Michealis-Menten like scheme where we include hydrolysis as a side reaction to show that the hydrolysis can help an enzyme to discriminate between two analogous substrates in spite of having similar binding energies.

The elementary steps of enzyme catalysis involve formation of Michaelis-Menten complex, followed by product formation. In recent years several experimental studies of enzyme catalysis in single molecular level revealed many interesting features in short time dynamics [11]. It successfully reproduces the same steady state rate as given by Michaelis-Menten kinetics. In single molecular enzyme catalysis there is only single enzyme present in the system and a steady flow of substrate is maintained as the concentration of substrate is relatively high with respect to the concentration of enzyme. In an in vitro experimental set up, the concentration of enzyme remains constant by means of recycling of the free enzyme after product release whereas the concentration of substrate decreases with time. In vivo situation can be different where concentration of enzyme can be similar to the concentration of substrate. In the present work we consider both the single molecular enzyme and the ensemble enzyme catalysis reactions.

In this article we mainly concentrate on amino-acylation of $tRNA^{Ile}$ with isoleucine as the correct and valine as the wrong substrate in presence of isoleucyl-tRNA synthetase (IRS) enzyme in the course of protein synthesis. This system has been extensively studied in several experiments [9, 12] and there are many interesting findings (mainly kinetic results) which are awaited to be resolved. For example, relative rates of amino-acylation of valine and isoleucine depends strongly on temperature. The proofreading by $\sim 10^4$ is not obvious from the difference in hydrolysis rate alone.

On the other hand, while a limited number of theoretical studies exist in the literature, they are mostly unable to explain the existing experimental results without any contradiction. In recent years due to the development of advance techniques in molecular spectroscopy there have been enormous attempts to analyze the structures of different intermediates formed in the course of reaction and enzyme [13, 14]. Though the structure of the various intermediates formed during the reaction can reveal the mechanism, it cannot give us the quantitative estimate of discrimination. Moreover it is often extremely difficult to obtain all the intermediates as many of them are transient. Kinetic experimental studies have the advantage of quantitative estimate of rate of a single step and also of the overall reaction. Therefore, the mechanism of a reaction can be understood by building up a reaction scheme based on the kinetic experimental data that can explain all the available kinetic results. The best way of understanding a reaction both quantitatively and qualitatively is to propose a scheme based on both of the kinetic and spectroscopic results and verify the experimental observations with the results derived from the scheme.

The structural analysis of tRNA aminoacylation enzymes reveals that there is an active site that activates amino acid in the presence of ATP and then the activated amino acid is transferred to the tRNA. The uniqueness of isoleucyl-tRNA synthetase (IRS) is that other than this active site it has separate editing site situated at $30A^0$ apart from the activation site in the domain known as CP1 domain (Fig.1) [13, 15-17]. The role of editing site is to hydrolyze the wrong enzyme-product complex leading to the lowering of error rate. The presence of this editing site makes IRS to discriminate valine more efficiently than that observed in non-editing tRNA-aminoacylation enzymes [18].

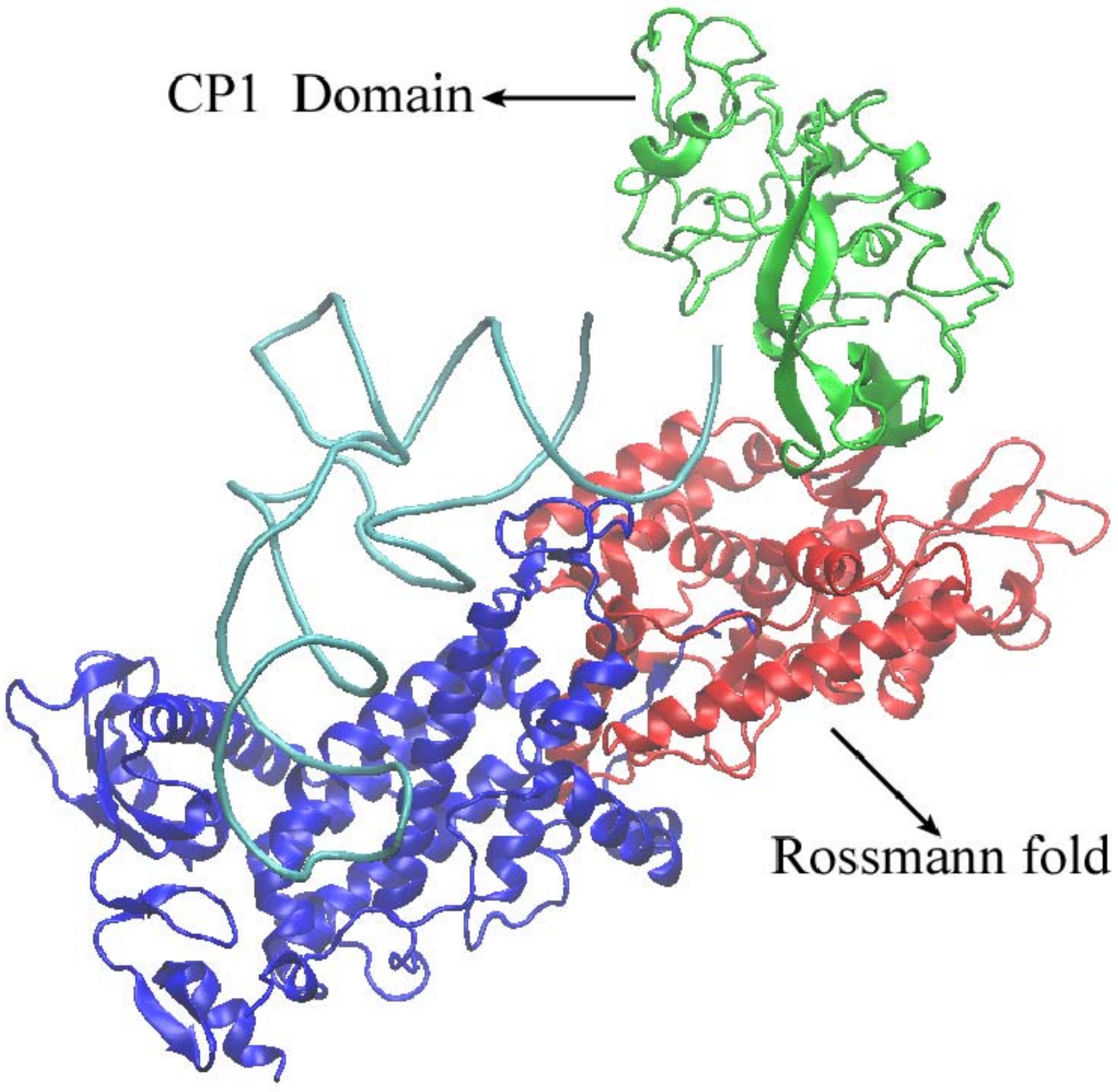

**Figure 1: Crystal structure of staphylococcus aureus IRS-tRNA complex (Protein Data Bank) [16]. tRNA is shown in magenta, CP1 domain in green, Rossmann flod in red, and rest of the protein is represented in blue.**

Monitoring the rate of ATP hydrolysis is important as this is directly related to the rate of hydrolysis of the intermediates formed during the reaction. The higher rate of ATP hydrolysis indicates higher rate of hydrolysis of the intermediates that serves as a side reaction giving rise to less amount of product formation. If there is no hydrolysis of the intermediates then the number of ATP molecule hydrolyzed per molecule of product formed would be identically 1. The value of the ratio in case of isoleucine is ~1.5 whereas in case of valine is ~270. Thus, ATP hydrolysis serves as a marker to gauge the accuracy of the proposed kinetic scheme.

Comparison of apo *Thermus thermophilus* LeuRS to the cocrystal structure of the aminoacylation complex of Pyrococcus horikoshii LeuRS with tRNALeu shows that the CP1 domain swings ~20° relative to the canonical core of the enzyme to prevent a clash with the bound 5′ terminus of tRNA [15]. Subsequent to aminoacylation, the 3′ end of the tRNA is translocated ~30 Å from the synthetic active site to the amino acid editing site for proofreading, requiring the editing domain to rotate by ~35° compared to the apo state of T. *thermophilus* LeuRS.

The accumulation of numerous LeuRS crystal structures shows clearly that the enzyme undergoes significant conformational changes to accommodate the aminoacylation and the editing complexes. This is also accompanied by changes in the tRNA, particularly at the 3′ end that is charged, and its interactions with the AARS.

The cocrystal of S. aureus IleRS with tRNAIle showed that the editing site within CP1 can be accessed by the 3′ end of the bound tRNAIle. A conformational change in the bound tRNA was proposed to shuttle the 3′ end from the active site to the editing site.

An elegant analysis of kinetic proofreading explaining enhanced specificity of phosphorylation-dephosphorylation cycle (PdPC) was carried out recently by Qian [19]. PdPC exhibits zero order ultra-sensitivity and known to exhibit sensitivity amplification. PDPC can also discriminate against non-specific cross talk in signal transduction. In fact, many kinetic proofreading mechanisms depende on a feedback mechanism that gives rise to a switch that describes the amplification of sensitivity.

## II. Proposed new kinetic scheme and method of solution

In the Fersht's alternative scheme [9], repeated activation is not required. Here the critical step is the enhanced rate of hydrolysis of the wrong substrate. Fersht described two mechanisms as follows:

**Fersht mechanism I**

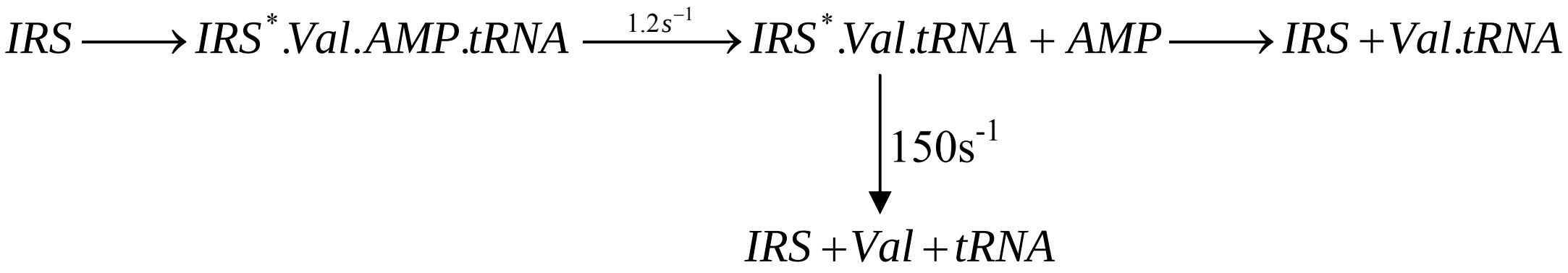

Here Val.AMP complex forms after hydrolysis of ATP, and this complex is bound to enzyme IRS. The rate constants are discussed later.

**Fersht mechanism II**

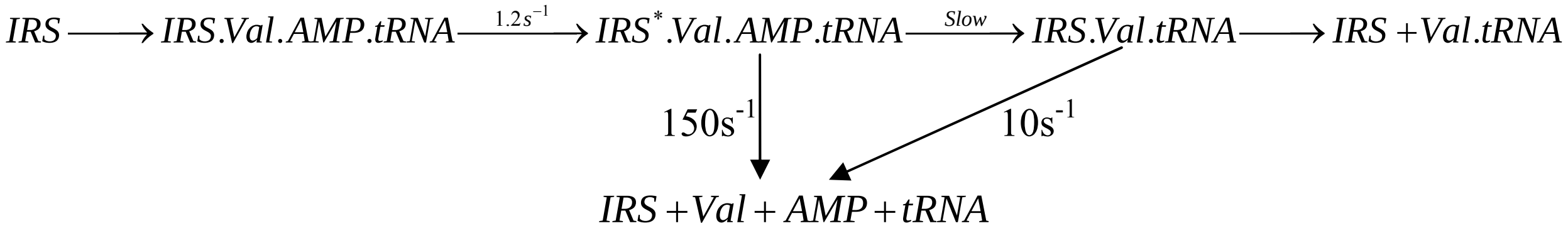

According to mechanism I, in case of valine the enzyme goes to a highly hydrolysable state during the formation of IRS$^*$.Val-AMP.tRNA$^{Ile}$ complex. After the rate determining step *i.e.*, the transfer of valine to $tRNA^{Ile}$ the complex gets rapidly hydrolyzed leading to the free enzyme and original substrate. Thus the hydrolysis is proposed to increase the discrimination from 10 to ~180. In an alternative mechanism (mechanism II) Fersht proposed that there is a possibility of the enzyme to undergo conformational fluctuation and go to highly hydrolysable conformation even before the transfer of valine to $tRNA^{Ile}$ (forming $IRS^{*}$.Val-AMP.$tRNA^{Ile}$) and that is the rate limiting step. After the conformational fluctuation, the Val-AMP.$tRNA^{Ile}$ complex gets rapidly hydrolyzed. There is another hydrolysis step involved with the product (Val-tRNA) if formed by mistake. On the basis of the experimental results, Fersht concluded that both of these two mechanisms are equally probable.

However, there are some difficulties with both the mechanisms. In both the cases it is assumed that the enzyme undergoes conformation change to a highly hydrolysable state. Structural analysis of IRS shows that it has an editing site situated at CP1 domain which is ~30$A^0$ apart from the active site (see Fig.1) and valine is translocated to this site after transfer to the $tRNA^{Ile}$ and gets hydrolyzed easily [13, 20-26]. This suggests that the enzyme does not need to undergo any conformational fluctuation. Rather, there is a binding site within the enzyme and the substrate goes to that site and gets hydrolyzed. As this translocation occurs after formation of Val-$tRNA^{Ile}$ bond, the hydrolysis step is involved after the formation of Val-$tRNA^{Ile}$ as in mechanism I. This point argues against mechanism II.

Now, according to mechanism I, the enzyme goes to the hydrolysable state during formation of Val-AMP.$IRS^{*}$.$tRNA^{Ile}$ complex and it gets hydrolyzed after formation of Val-$tRNA^{Ile}$ bond. The rate determining step is considered to be the step before hydrolysis *i.e.*, the step which involves the bond formation between amino acid and tRNA. However, several experimental evidences confirm that the product release step is the rate limiting step.

Thus, none of the above mechanisms take care of the hydrolysis properly. Here we construct a mechanism based both on kinetic experiments and structural evidences and carry out quantitative study to understand the contribution of the enhanced hydrolysis in overall discrimination and ATP hydrolysis. We point out that such a quantitative analysis has not been carried out even for Fersht mechanisms.

Based on Fersht and Hopfield's experimental results, we construct a new theoretical model that has the following features.

i. The hydrolysis occurs only after enzyme bound product formation but before product release.

ii. Enzyme-substrate complex (ES) formation step involves a small discrimination (~1/10) [27] as in Hopefield's scheme. But the reaction of this complex with ATP involves further discrepancy leading to overall discrimination of 1/160 up to the formation of enzyme-substrate-AMP complex.

iii. In case of wrong substrate (valine) transfer to tRNA, the amino acid goes to the CP1 domain (hydrolytically active region) of the enzyme. As a result it undergoes rapid dissociation through enhanced hydrolysis.

iv. The extremely low concentration of enzyme-product complex at steady state, as found by Fersht in case of valine, suggests that it undergoes enhanced hydrolysis

Based on above observations, we developed a new scheme which is given (for isoleucine) as follows.

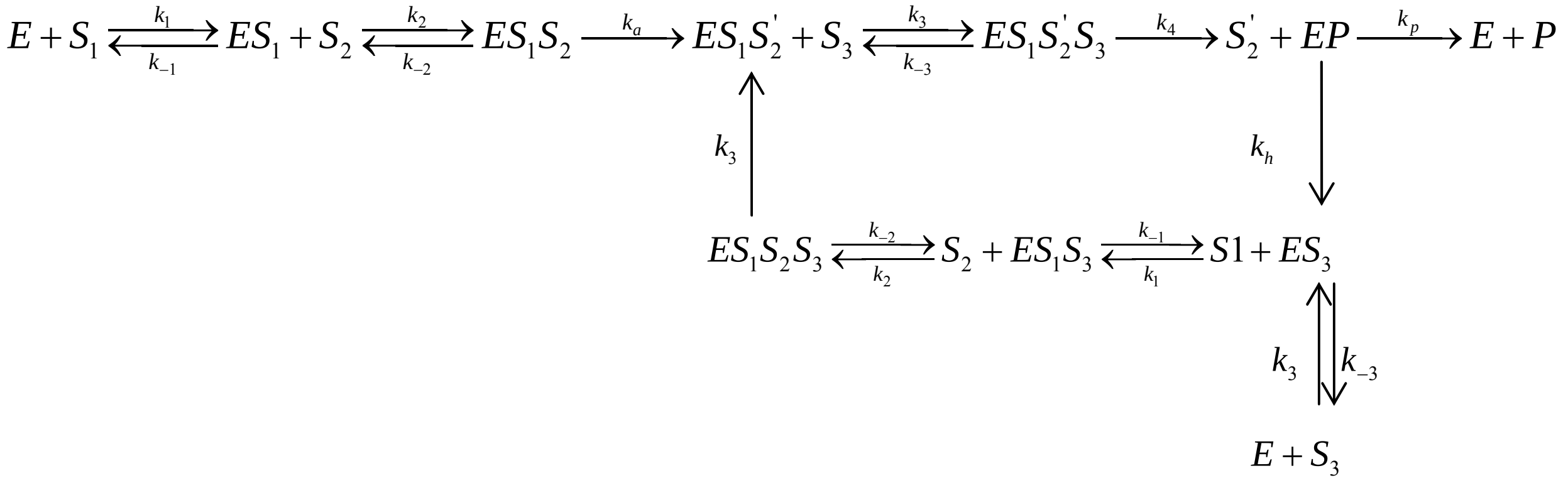

The proposed modified reaction scheme of valine is a bit different and given as follows:

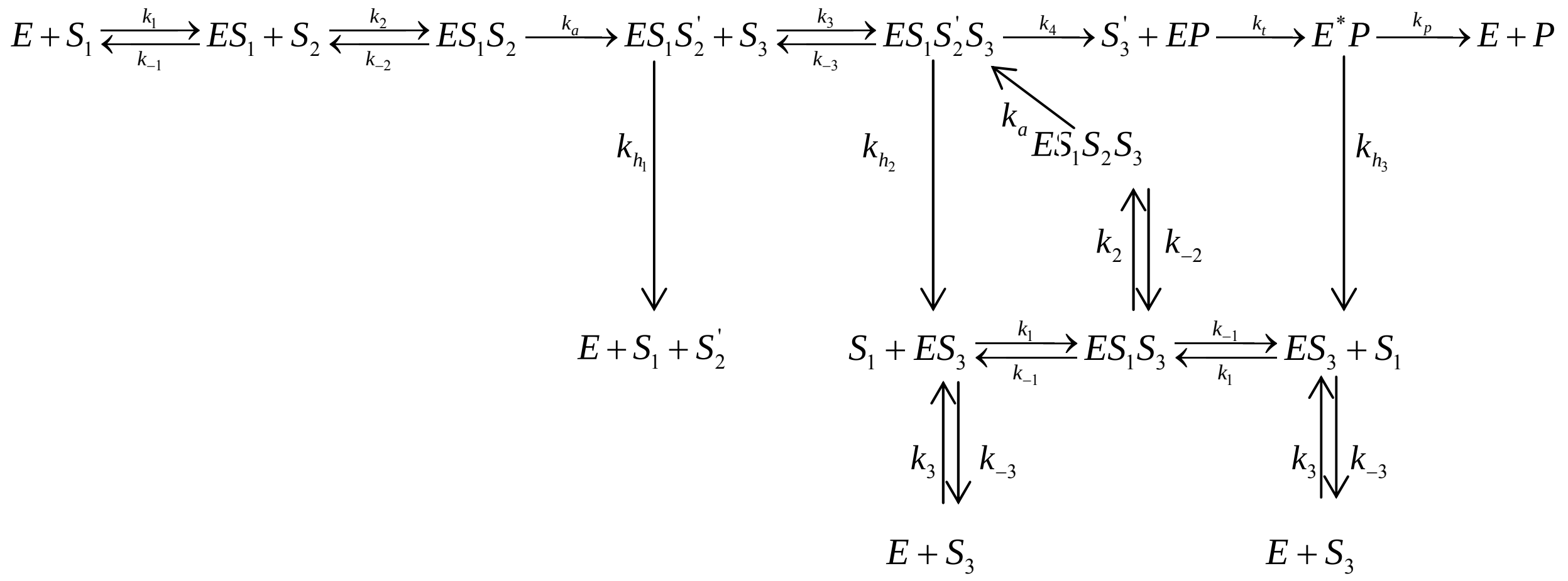

Here, $E$ is IRS, $S_1$ is amino acid, $S_2$ is ATP, $S_2^{'}$ is AMP, $S_3$ is tRNA$^{\text{Ile}}$, $P$ is the product (either $Ile-tRNA^{Ile}$ or $Val-tRNA^{Ile}$), $E^*$ indicates that the product is in CP1 domain.

We now determine the time dependent rate of catalysis and the concentrations of ATP hydrolysis for both substrates. The above kinetic schemes can be represented by a set of simple chemical kinetic equations. The solution of these equations provides the time dependent concentration of substrates, different intermediates and products from which the rate of catalysis and ATP hydrolysis can be obtained easily. Since the above schemes are complicated the analytical solution is too difficult to obtain. Though one can find a solution for a given set of kinetic parameters following numerical integration method, the solution obtained by this method is deterministic in nature. On the other hand, the in vivo reactions are non-deterministic because of small size of biological cell. There is an alternative method of obtaining the non-deterministic solution following the method of first passage time distribution. This method follows a probabilistic approach a finds the first passage time distribution of product formation from which the steady state rate can easily be obtained. Two main limitations of this approach are that for complicated reactions that involves various elementary steps such as above kinetic schemes it fails. The other and the most important drawback is that it does not provide the time dependence concentrations of substrates, intermediates and products since it deals with the probability of first passage time not probabilities of the substances.

In the present study we have carried out a stochastic simulation analysis proposed by Gillespie and obtained time dependent concentrations of all the substrates, intermediates and products [28]. This is a nondeterministic method and gives concentrations of various substances with progress of time. Whereas for small system it shows randomness in concentration along time, a deterministic time-dependent concentration is obtained for large system. Moreover, both the single molecular as well as ensemble enzyme catalysis can be studied following this method.

## III. Results and discussions

The above kinetic schemes translate into a system of kinetic equations familiar from chemical kinetics, particularly in the area of enzyme kinetics. There are several methods to solve such coupled system of first order differential equations. One approach we adopted earlier is the method of first passage time distribution (f(t)) by directly deriving an equation for f(t).. This method is practical when the number of coupled equations is small, like six or less. Here however we have to deal with a large number of concentrations, more than ten. So, we used the method of stochastic simulation method of Gillespie to obtain the time dependent concentrations [28]. This method and its different variations have been discussed extensively in the literature.

In order to carry out the analysis we need numerical values of all the kinetic parameters. We have considered the following set of k's for isoleucine that are either obtained or derived from literature.

$k_1 = 2.5\times10^7\, M^{-1}s^{-1}, k_{-1} = 72.5\, s^{-1}, k_2 = 4.8\times10^4\, M^{-1}s^{-1}, k_{-2} = 2\, s^{-1}, k_a = 28\, s^{-1}$,

$k_3 = 5.5\times10^7\, M^{-1}s^{-1}, k_{-3} = 10\, s^{-1}, k_r = 15\, s^{-1}, k_h = 0.048\, s^{-1}$ and $k_p = 1.2\, s^{-1}$.

Similarly, the rate constants for valine are,

$k_1 = 4.1\times10^4\, M^{-1}s^{-1}, k_{-1} = 1.4\, s^{-1}, k_2 = 4.8\times10^4\, M^{-1}s^{-1}, k_{-2} = 2\, s^{-1}, k_a = 17\, s^{-1}$, $k_3 = 5.5\times10^7\, M^{-1}s^{-1}$,

$k_{-3} = 10\, s^{-1}$, $k_r = 15\, s^{-1}$, $k_t = 1.2\, s^{-1}, k_{h_1} = 0.053\, s^{-1}, k_{h_2} = 1.2\, s^{-1}, k_{h_3} = 150\, s^{-1}$ and $k_p = 0.85\, s^{-1}$.

First, we have studied the single molecular catalysis for both the substrates. In this study a single enzyme is considered and the other substrates (amino acids, ATP and $tRNA^{Ile}$) are taken in excess amount which is similar to the single molecular experimental methodology. Fig.2 and Fig.3 show the waiting time distributions for the Ile and Val, respectively. Figures clearly show that the time scales of product formation for isolucine and valine are vastly different. The rate of product formation is identical to the first passage time distribution. The first passage time distributions show that the rate of product formation for isoleucine decays very fast and goes to zero within a second. On the other hand the corresponding rate for valine does not change much even up to 10 seconds (shown in inset of Fig. 3). Experimental study also shows the similar behavior [12]. The large difference in time scale for the two substrates arises due to the presence of pre-transfer and post-transfer hydrolysis in case of valine.

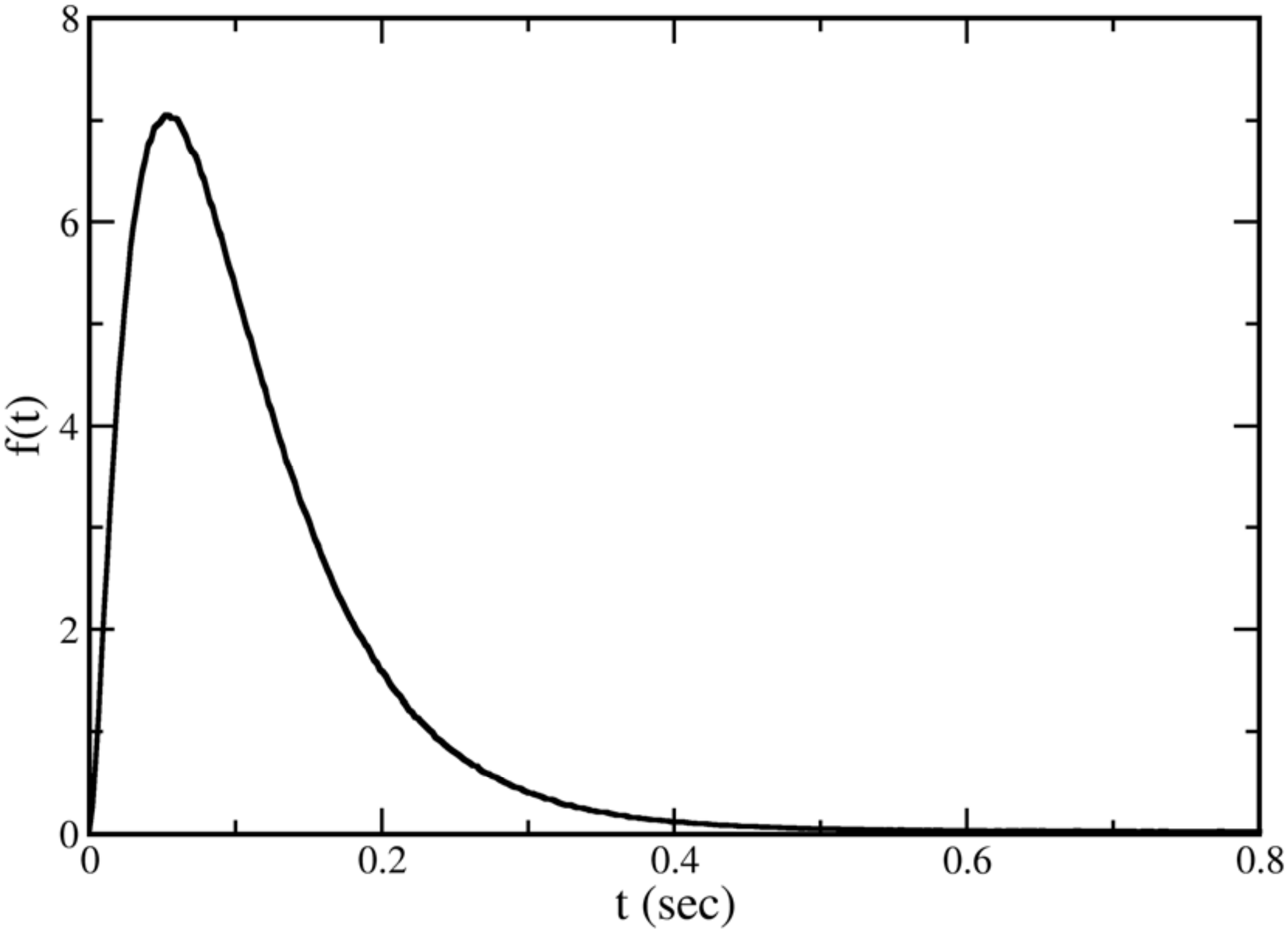

**Figure 2: The first passage time distribution for the aminoacylation of $tRNA^{Ile}$ by isoleucine in single molecular reaction, under single molecular catalytic conditions.**

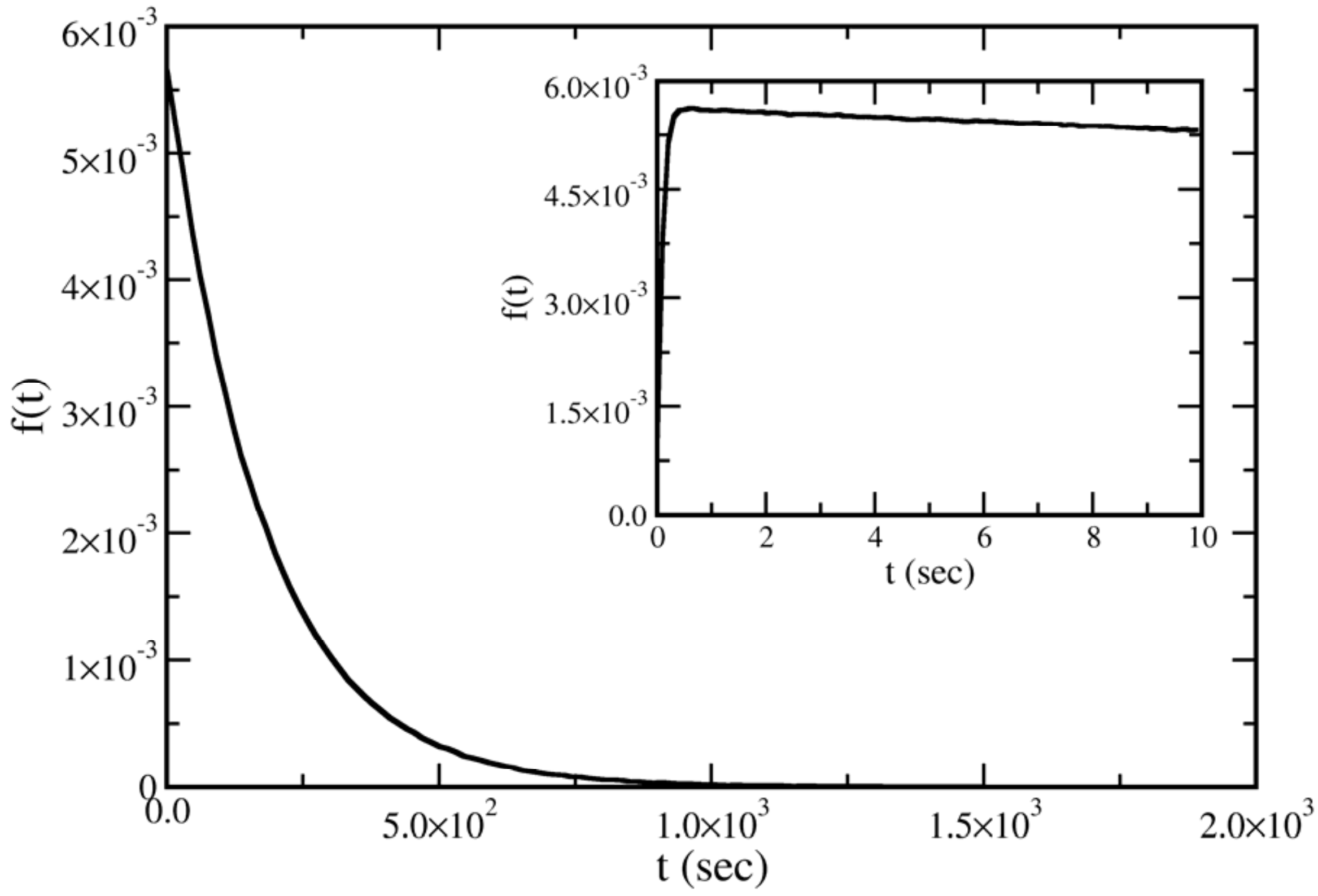

**Figure 3: The computed first passage time distribution for the aminoacylation of $tRNA^{Ile}$ by valine in single molecular reaction. The inset shows the time dependence at short time where the distribution shows an initial sharp rise followed by a slow decay.**

The role of $tRNA^{Ile}$ on the hydrolysis of ATP for valine is presented in Fig. 4. It shows that the ATP hydrolysis rate in steady state increases with increasing concentration of $tRNA^{Ile}$. At low $tRNA^{Ile}$ concentration it increases fast and then it saturates at high concentration. Similar behavior is observed in case isoleucine (data not shown). The steady state rate of ATP hydrolysis for both valine (Fig. 5(A)) and isoleucine (Fig. 5(B)) shows Michaelis-Menten type behavior. Interestingly, the inverse rate versus inverse $tRNA^{Ile}$ concentration plot in case of valine (inset of Fig. 5(A)) indicates a deviation from Michaelis-Menten behavior incontrast to the isoleucine (inset of Fig. 5(B)). This deviation is attributed to the correlation of relative contribution of tRNA-independent hydrolysis to tRNA-dependent hydrolysis. Thus, the presence of tRNA-independent hydrolysis makes the dependency of overall ATP hydrolysis on $tRNA^{Ile}$ complicated.

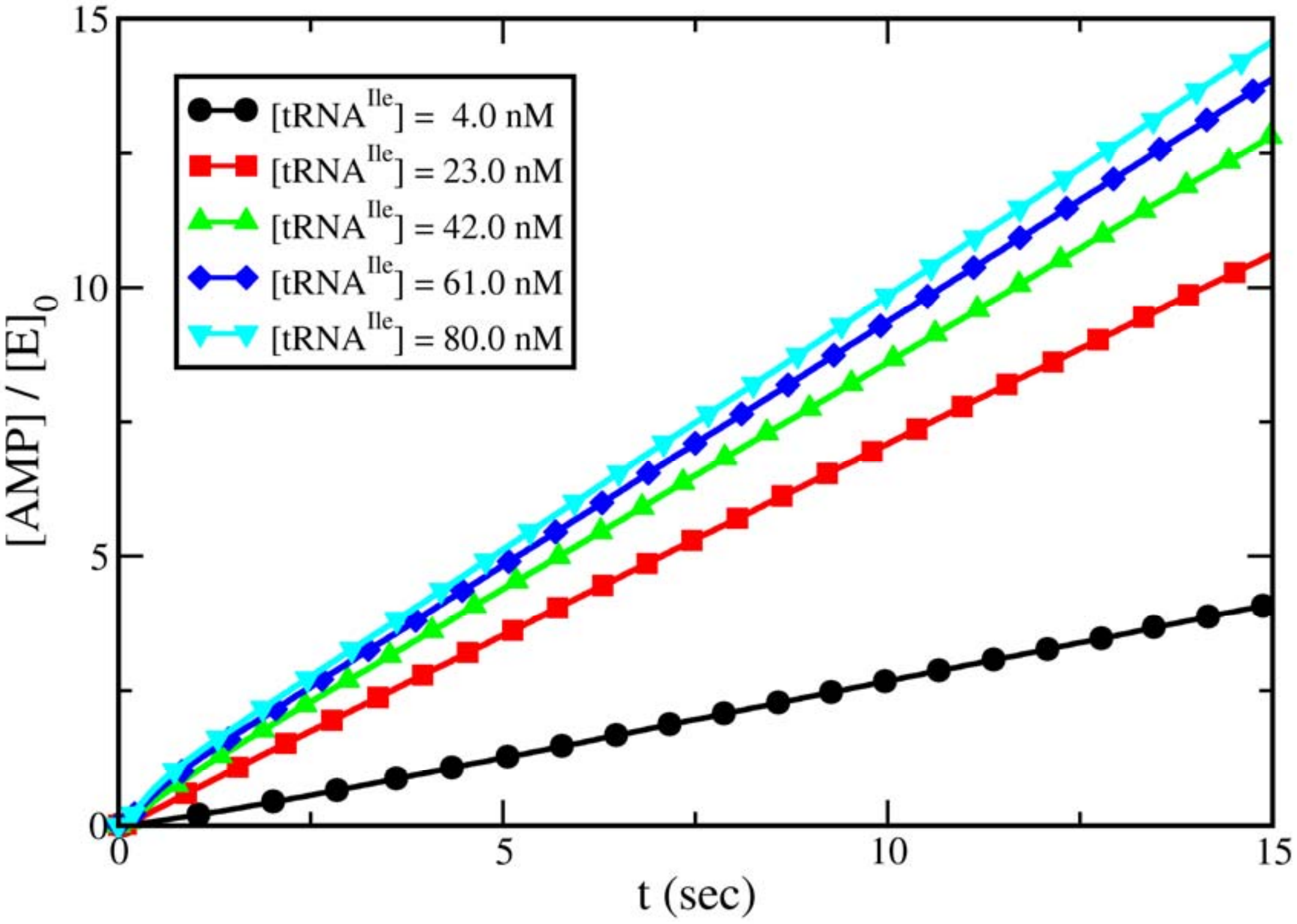

**Figure 4: Time dependence of AMP concentration generated by ATP hydrolysis for different $tRNA^{Ile}$ concentrations during synthesis of Val-$tRNA^{Ile}$.**

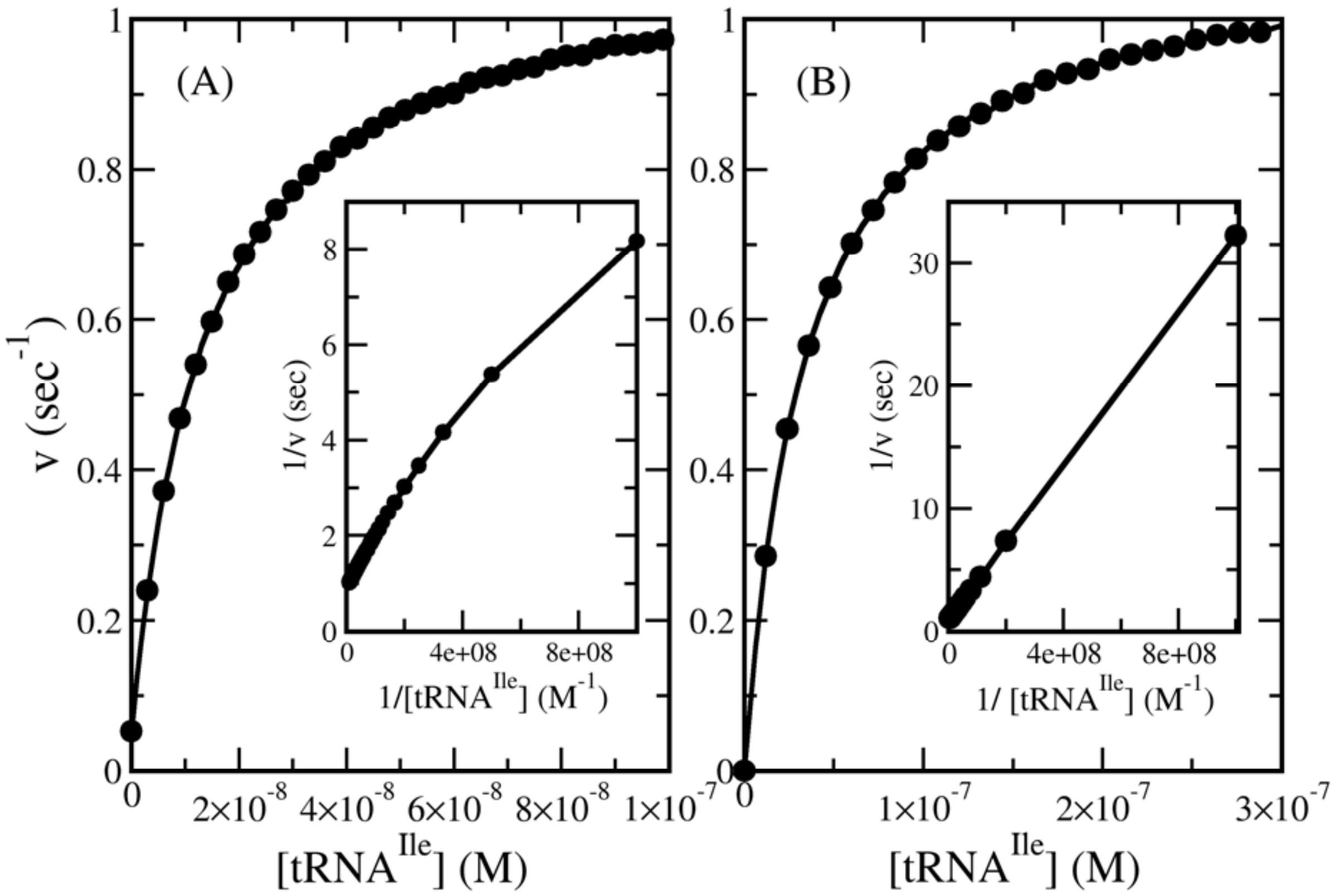

**Figure 5: tRNA$^{Ile}$ concentration dependence of steady state rate of ATP hydrolysis for (A) valine and (B) isoleucine. The insets show their corresponding behaviors in inverse scales. Note the deviation from Michaelis-Menten behavior for isoleucine.**

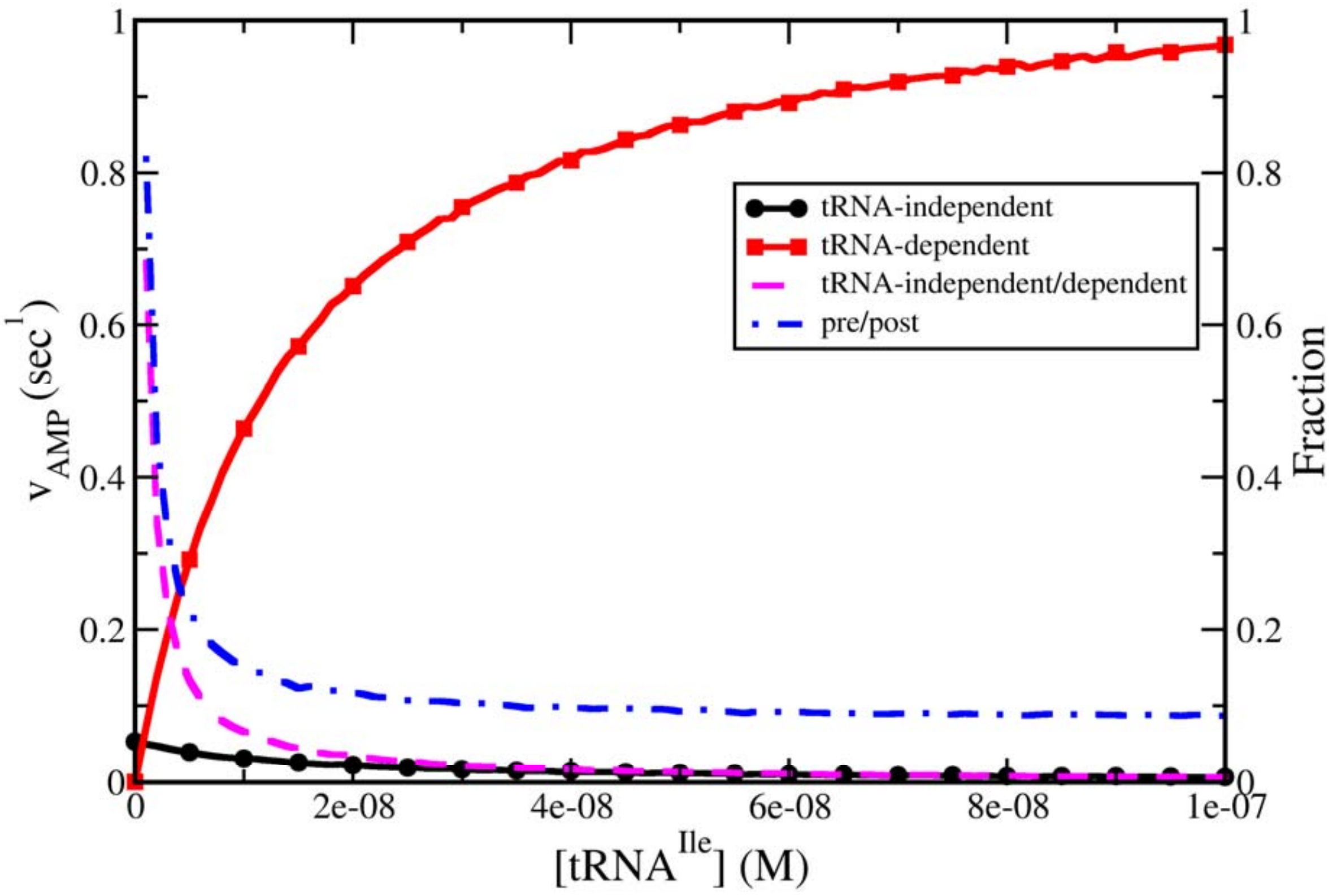

**Figure 6: tRNA$^{Ile}$ concentration dependent steady state rate of ATP hydrolysis for Val in pre- and post transfer processes. The fraction of pre-transfer hydrolysis with respect to the overall ATP hydrolysis is also shown with a blue dash line. The relative contribution of pre-transfer editing converges to a finite value (~9%) at high tRNA$^{Ile}$ concentration.**

Most of the recent experimental studies on kinetic proofreading of tRNA-aminoacylation try to understand the relative contributions of pre-transfer and post-transfer editing as well as the tRNA-independent and dependent editing, since the mechanism of editing is reflected on these quantities. A minor contribution of tRNA-independent editing through hydrolysis for valine has been observed repeatedly which is indicated by the rate of ATP hydrolysis in absence of tRNA$^{Ile}$. However, experimentally it is not possible in to estimate the relative contribution of tRNA-independent editing in presence of $tRNA^{Ile}$. In our present study we have tagged the AMP molecules that are produced by hydrolysis in tRNA-independent editing step. Both tRNA-dependent and independent rate of ATP hydrolysis are presented in Fig. 6. As expected the rate of tRNA-dependent ATP hydrolysis increases with increase of $tRNA^{Ile}$ concentration. Intersetingly, the rate of tRNA-independent ATP hydrolysis decreases with increase of $tRNA^{Ile}$ concentration and decays to zero at high concentration. It is to be noted that though the tRNA-independent ATP hydrolysis goes to zero, the relative contribution of pre-transfer editing decreases and then converges to a finite value (~9%) at high $tRNA^{Ile}$ concentration.

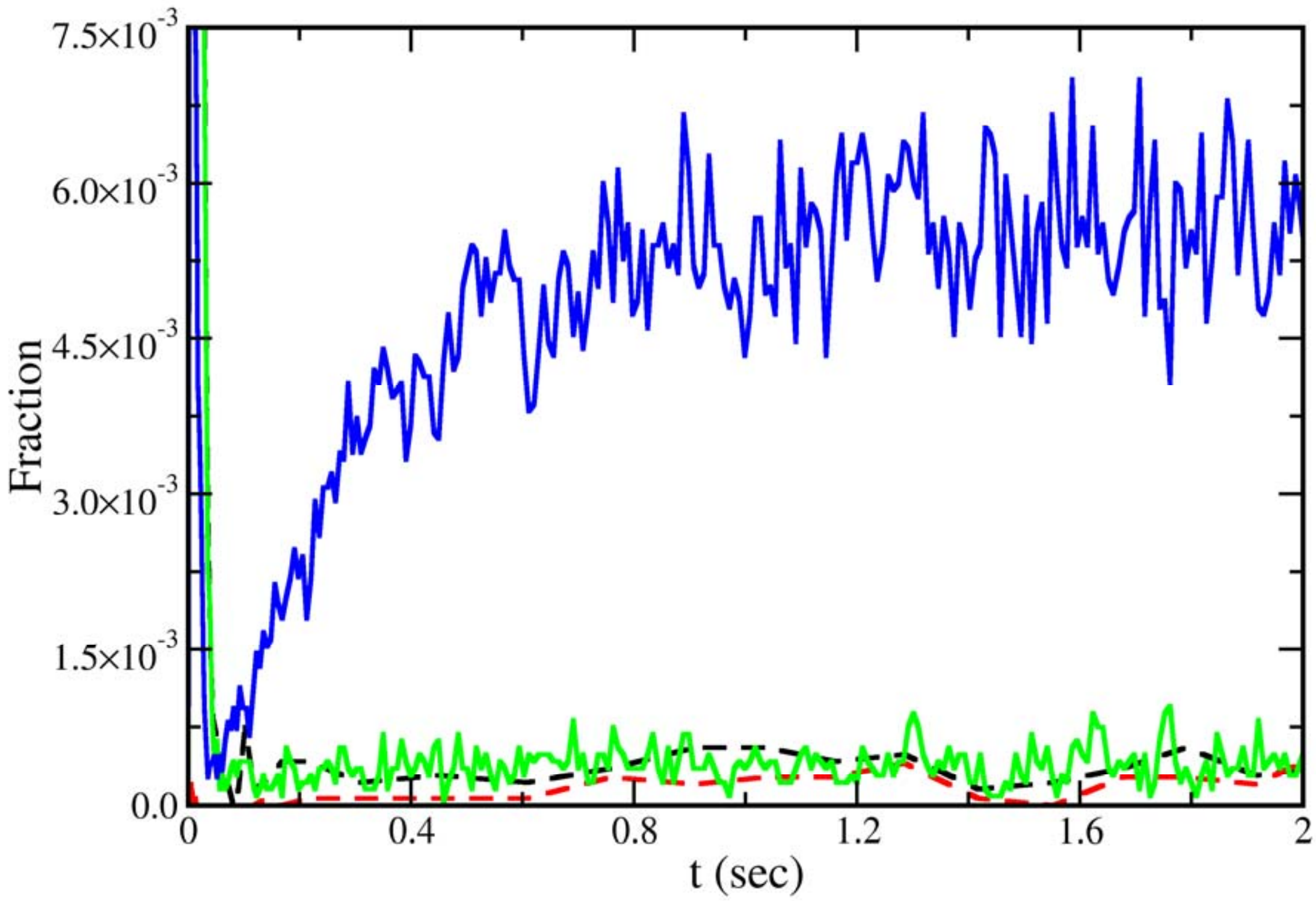

**Figure 7: Time dependence of fraction of free enzyme and enzyme-$tRNA^{Ile}$ complex at two different $tRNA^{Ile}$ concentrations ([$tRNA^{Ile}$]=1nM and 100nM) under multi-turnover condition in presence of valine. The dashed and solid lines are for 1nM and 100nM**

**concentrations of tRNA$^{Ile}$, respectively. Black and green lines indicate fraction of free enzyme whereas the red and blue lines indicate fraction of enzyme-tRNA complex.**

The decreases of relative contribution of tRNA-independent editing is because of the fact that the population of enzyme-tRNA$^{Ile}$ increases and free enzyme decreases with increasing tRNA$^{Ile}$ concentration. At very high tRNA$^{Ile}$ concentration almost all the enzymes remain as enzyme-tRNA$^{Ile}$ complex whereas the tRNA-independent editing only occurs when enzyme starts catalytic cycle form free form. In Fig. 7 we show the fraction of free enzyme and enzyme-tRNA$^{Ile}$ complex at two limiting concentrations of tRNA$^{Ile}$. At low concentration enzymes remain in both forms. On the other hand, at high concentration it shows a significant amount of enzyme-tRNA$^{Ile}$ complex compared to free enzyme.

The steady state rates of catalysis for isoleucine and valine and ATP hydrolysis for valine are shown in Fig. 8. The amino acid concentration dependence of reaction rates for both the substrates depicts Michaelis-Menten like behavior. Although the rate of ATP hydrolysis for valine is different from the rate of reaction for isoleucine at small to intermediate amino acid concentration, they are almost equal at high concentration of amino acids and saturate to the value of ~$1.1 s^{-1}$.

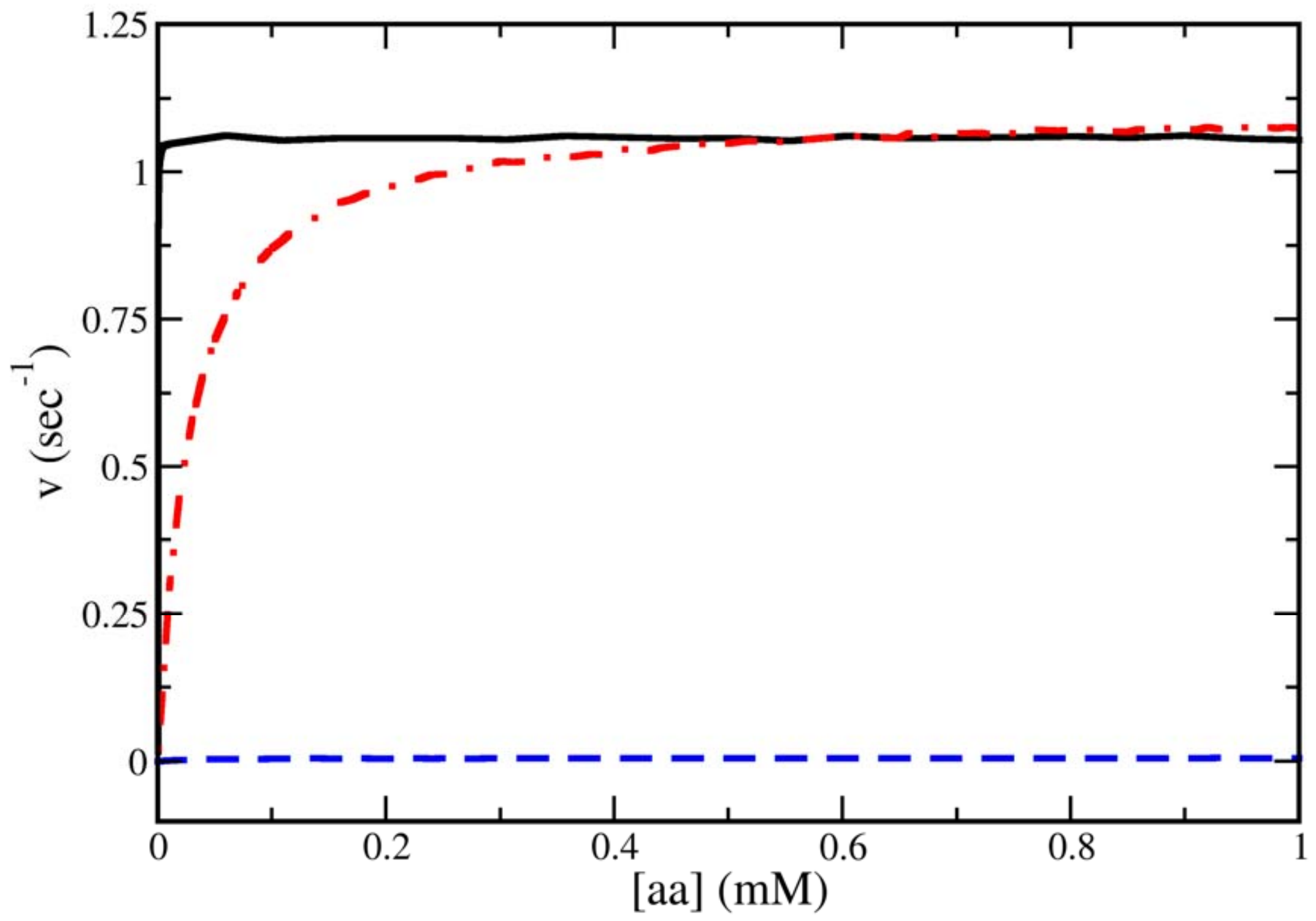

**Figure 8: Amino acid concentration dependence of steady state rate of enzyme catalysis for isoleucine (black solid line) and valine (blue dashed line). The rate of ATP hydrolysis for Val is shown by red dot-dashed line.**

The rate of amino-acylation of isoleucine was found to be higher than the rate of ATP hydrolysis in case of valine at intermediate concentration (72 μM) of substrate [12]. At very high concentration of substrate a different experimental study shows that these two rates are exactly same at various temperatures and pH of the system [9]. It is important to note that though the rate of aminoacylation of isoleucine is same as the rate of ATP hydrolysis in case of valine at high substrate concentration, the former rate is always higher than the latter at low to intermediate amino acid concentration. From the amino acid concentration dependence of steady state rate of product formation for isoleucine and valine we find that the michaelis constants are ~60 nM and ~26 μM, respectively. As a result, in micromolar concentration range the steady state rate of product formation of isoleucine becomes saturated but, the ATP hydrolysis for valine still remains below its maximum value. The difference in scale of michaelis constants of isoleucine and valine makes the ATP hysrolysis rate for valine to be less compared to the rate of product formation rate of isoleucine at intermediate concentration range (upto 500 μM). The rate of product formation of valine remains low throughout the amino acid concentration range (Fig. 8).

At this point it is necessary to mention that the steady state concentration of Val-$tRNA^{Ile}$ is less than 0.8% of either the IRS or the IRS.Val-AMP complex. [9] Both the high value of the ATP to product ratio and the low concentration of Val-$tRNA^{Ile}$ suggest that *the rate of hydrolysis during the reaction is much higher than the normal hydrolysis rate of Val-$tRNA^{Ile}$*. This is possible if the amino acid goes to the highly hydrolyzing CP1 domain of IRS after getting transferred to the $tRNA^{Ile}$. Inclusion of such enhanced hydrolysis during the reaction explains both the high value of the ratio and the low concentration of Val-$tRNA^{Ile}$.

The overall discrimination when both the substrates present in the reaction mixture with equal amount is the ratio of $k_{cat}/K_m$ for wrong and correct substrates. Thus, the discrimination, $D = \left(\frac{k_{cat}}{K_m}\right)_{Ile} \Big/ \left(\frac{k_{cat}}{K_m}\right)_{Val}$. In our present study we obtained $(k_{cat})_{Ile} = 1.057 s^{-1}$ and $(k_{cat})_{Val} = 5.76\times10^{-3}\ s^{-1}$. Thus, the ratio $(k_{cat})_{Ile}/(k_{cat})_{Val}$ is $1.8\times10^2$. We also obtained the values of $(K_m)_{Ile}$ and $(K_m)_{Val}$ and these are $6.17\times10^{-8}$ M and $2.68\times10^{-5}$ M, respectively. Then the ratio $(K_m)_{Ile}/(K_m)_{Val}$ is obtained to be $4.33\times10^2$. It should be noted here that the experimentally determined value of $(K_m)_{Ile}/(K_m)_{Val}$ is $\sim1/10^2$ [27]. Therefore, the overall discrimination is $(1.8\times10^2).\ (4.3\times10^2) = 7.8\times10^4$.

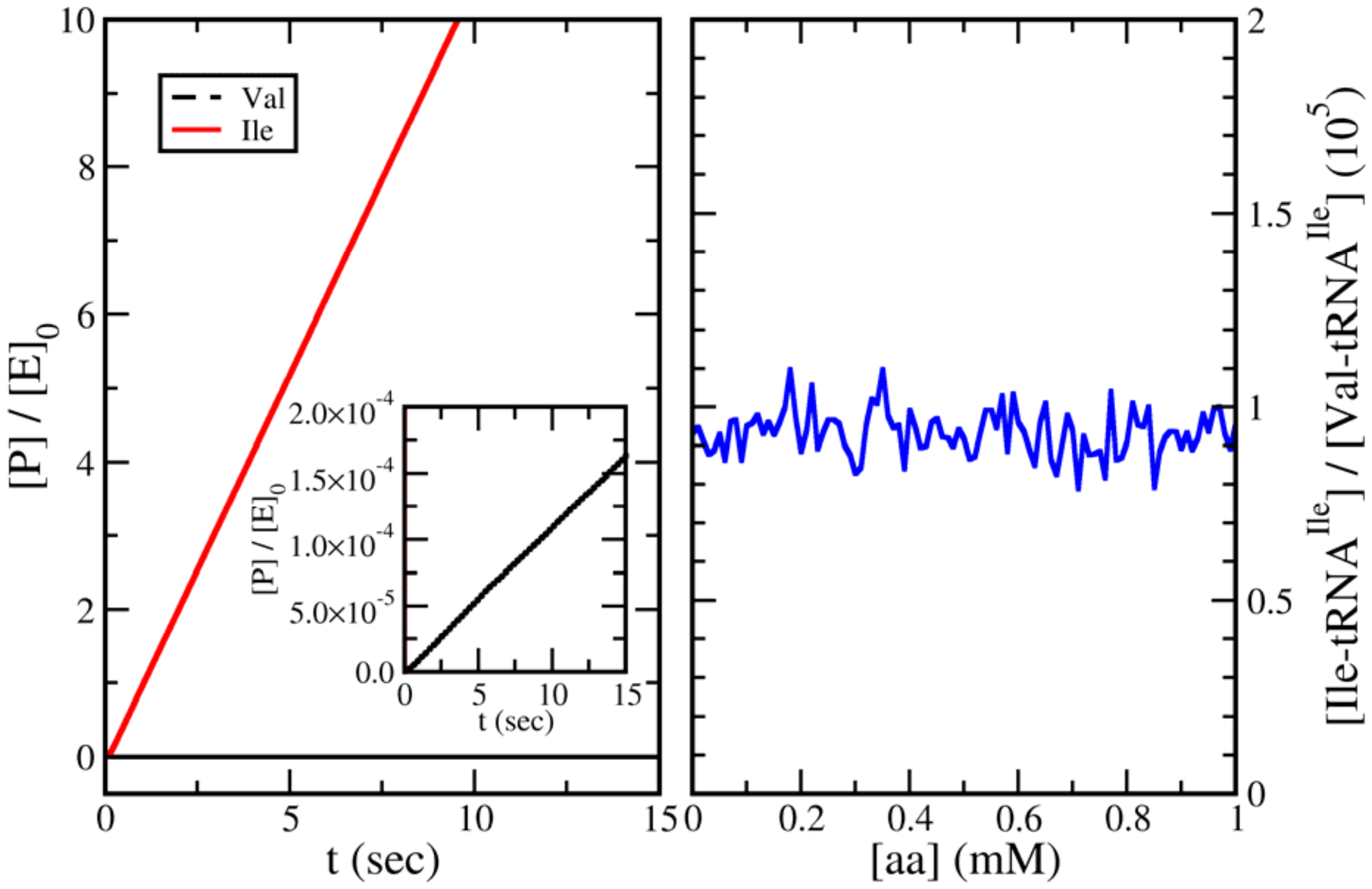

**Figure 9: (A) The time dependence of products (Ile-tRNA$^{Ile}$ and Val-tRNA$^{Ile}$) when both isoleucine and valine are present in the system with equal amount. Inset shows the linear increase of Val-tRNA$^{Ile}$ during steady state. (B) Amino acid concentration dependence of overall editing (i.e., the ratio of steady state rates of Ile-tRNA$^{Ile}$ and Val-tRNA$^{Ile}$ formation in presence of equal amounts of Ile and Val).**

So far we have presented the kinetics of isoleucine and valine and corresponding editing treating the substrates (isoleucine and valine) separately. To obtain the overall editing directly we perform analysis in presence both of the substrates with equal amount at saturating condition of tRNA$^{Ile}$ and ATP. The time-dependent products are sown in Fig. 9(A). The figure and corresponding inset indicate that the overall editing is ~$10^5$ which is close to the above mentioned value of D. We find that the editing in steady state does not dependent on the concentration of amino acids present in the system (Fig. 9(B)).

## IV. Conclusion

Given the multitude of processes involved in aminoacylation of tRNA, a quantitative understanding of kinetic proofreading is not easy. The existence of many rate constants makes accurate experimental elucidation of the proofreading mechanism also difficult. Thus, the goal is not just to differentiate between Hopfield and Fersht schemes, but also to rationalize the rate constants observed experimentally.

Advances of single molecule spectroscopy offers hope as one can obtain the waiting time distribution which contains more information than just the steady state rate. Also, there is the possibility that we can observe individual processes separately and individually.

In this article we have introduced a modified Fersht scheme to quantitatively understand the origin of efficient kinetic proofreading in the aminoacylation of $tRNA^{Ile}$ by considering the specific case of discrimination between isoleucine and valine. Fersht scheme proposes hydrolysis as the key discrimination step. With our proposed scheme we have obtained the first passage time distribution as well as the steady state rate of reaction and ATP hydrolysis for Ile and Val. The present theoretical study successfully capture the experimental observations such that the difference in rate of overall reaction. We show that the steady state rate of ATP hydrolysis at high Val concentration is same as the rate of product formation for Ile at its high concentration whereas at low to intermediate concentration of amino acids ATP hydrolysis for valine is less than the rate of Ile-tRNA$^{Ile}$ formation. The overall steady state rate of reaction in case of valine shows a non-michaelis type dependence on tRNA$^{Ile}$ concentration because of the interplay between tRNA-independent and tRNA-dependent editing. The contribution of tRNA-independent editing on overall editing for valine decreases with increasing tRNA$^{Ile}$ concentration and goes to zero at high concentration. We find that a significant amount of enzyme remains as enzyme-tRNA$^{Ile}$ complex compared to free enzyme and it lowers the relative contribution of tRNA-independent editing.

The enhanced hydrolysis of the wrong substrate is clearly a spontaneous and natural process, particularly due to the abundance of water in biological cells. The present work quantifies the importance of the hydrolysis step and finds that in a reasonable kinetic scheme, this step can broadly explain the magnitude of discrimination involved in kinetic proofreading.

There are still many critical steps involved in protein synthesis such as the kinetic proofreading involved in the selection of aminoacyl-tRNA in ribosome [29, 30] and the selection of correct base pair in DNA replication. Difficulty in formulation of a correct kinetic scheme for any of these problems is the non-availability of the rate constants for several critical steps. Thus, even when a correct scheme proposed, its validity cannot be easily confirmed. Further experimental studies of the individual kinetic steps are required.

Molecular mechanism of editing is yet to be understood at a microscopic level. It is clear that CP1 and Rossmann fold catalytic domains can determine the course of the events subsequent to the product formation. It is possible that the size of the wrong product allows it to transfer to CP1 domain while the correct product gets released without going to CP1. Recent structureal analyses have revealed an evidence of large amplitude conformational motion of CP1 domain around glycine at the β-strand. Single molecular spectroscopic study can certainly help in elucidating the proper mechanism.

## Acknowledgement

We would like to thank Prof. Umesh Varshney for discussions and for pointing out valuable references. We thank Prof. Alan Fersht for helpful suggestions. MS thanks Mr. Rakesh Sharan Singh, Rajib Biswas and Saikat Banerjee for valuable inputs in preparing manuscript. This work was supported in parts by grants from DST and CSIR (India). BB thanks DST for a JC Bose Fellowship.

## References:

[1] Pauling, L. Festschrift Arthur Stoll Birhhauser (Basel) , 597 (1958).

[2] Lehninger, A. L. Worth Publishers, New York, (1970).

[3] Watson, J. D. W. A. Benjamin, New York, (1970).

[4] Kimura, M. and Ohta, T. Genetics 73, 19 (1973).

[5] Fersht, A. R. Proc. R. Soc. Lond. B 212, 351 (1981).

[6] Hopfield, J. J. Proc. Nat. Acad. Sci. USA 71(10), 4135 (1974).

[7] Hussain, T., Kamarthapu, V., Kruparani, S. P., Deshmukh, M. V., and Sankaranarayanan, R. Proc. Nat. Acad. Sci. USA 107, 22117 (2010).

[8] Boniecki, M. T., Vu, M. T., Betha, A. K., and Martinis, S. A. Proc. Nat. Acad. Sci. USA 105, 19223 (2008).

[9] Fersht, A. R. Biochemistry 16(5), 1025 (1977).\

[10] Santra M. and Bagchi B., J. Phys. Chem. B 116, 11809 (2012).

[11] Lu, H. P., Xun, L., and Xie, X. S. Science 282(5395), 1877–1882 (1998).

[12] Hopfield, J. J., Yamane, T., Yue, V., and Coutts, S. M. Proc. Nat. Acad. Sci. USA 73(4), 1164–1168 (1976).

[13] Fukai, S., Nureki, O., Sekine, S., Shimada, A., Tao, J., Vassylyev, D. G., and Yokoyma, S. Cell 103, 793 (2000).

[14] Mascarenhas, A. P. and Martinis, S. A. Biochemistry 47, 4808 (2008).

[15] Zhu, B., Yao, P., Tan, M., Eriani, G., and Wang, E. J. Biol. Chem. 284, 3418 (2009).

[16] Silvian, L. F., Wang, J., and Steitz, T. A. Science 285, 1074 (1999).

[17] Cusack, S., Yaremchuk, A., and Tukalo, M. EMBO J. 19, 2351 (2000).

[18] Gruic-Sovulj, I., Uter, N., Bullock, T., and Perona, J. J. J. Biol. Chem. 280, 23978 (2005).

[19] Qian H., Biophy. Chem. 105, 585 (2003).

[20] Splan, K. E., Ignatov, M. E., and Musier-Forsyth, K. J. Biol. Chem. 283, 7128 (2008).

[21] Apostol, I., Levine, J., Lippincott, J., Leach, J., Hess, E., Glascock, C. B., Weickert, M. J., and Blackmore, R. J. Biol. Chem. 272, 28980 (1997).

[22] Zhao, M. W., Zhu, B., Hao, R., Xu, M. G., Eriani, G., and Wang, E. D. EMBO J. 24, 1430 (2005).

[23] Fersht, A. Freeman, New York, 2nd edition, (1985).

[24] Eriani, G., Delarue, M., Poch, O., Gangloff, J., and Moras, D. Nature 347, 203 (1990).

[25] Nomanbhoy, T. K., Hendrickson, T. L., and Schimmel, P. Mol. Cell. 4, 519 (1999).

[26] Nomanbhoy, T. K. and Schimmel, P. R. Proc. Nat. Acad. Sci. USA 97, 5119 (2000).

[27] Loftfield, R. B. and Eigner, E. A. Biochim. Biophys. Acta. 130, 426 (1966).

[28] Gillespie D. T., J. Comput. Phys., 22, 403 (1976).

[29] Blanchard, S. C., Jr, R. L. G., Kim, J. D., Chu, S., and Puglisi, J. D. Nat. Struc. Mol. Biol. 11, 1008 (2004).

[30] Schmeing, T. M. and Ramakrishnan, V. Nature 461, 1234 (2009).